\documentclass[twocolumn,preprintnumbers,amsmath,amssymb]{revtex4}

\usepackage{graphicx}
\usepackage{dcolumn}
\usepackage{bm}

\newcommand{\nc}{\newcommand}

\nc{\ra}{\rightarrow}

\nc{\da}{\dagger}
\nc{\dn}{\downarrow}
\nc{\cA}{{\cal A}}
\nc{\cB}{{\cal B}}
\nc{\cC}{{\cal C}}
\nc{\cD}{{\cal D}}
\nc{\cE}{{\cal E}}
\nc{\cF}{{\cal F}}
\nc{\cG}{{\cal G}}
\nc{\cH}{{\cal H}}
\nc{\cI}{{\cal I}}
\nc{\cJ}{{\cal J}}
\nc{\cK}{{\cal K}}
\nc{\cL}{{\cal L}}
\nc{\cR}{{\cal R}}
\nc{\cS}{{\cal S}}
\nc{\cX}{{\cal X}}
\nc{\cO}{{\cal O}}
\nc{\cU}{{\cal U}}

\begin{document}

\title{Negative Entropy and Black Hole Information}

\author{D. Song}

\affiliation{School of Liberal Arts, KOREATECH, Chungnam 330-708, Korea}%

\date{\today}

\begin{abstract}

Based on negative entropy in entanglement, it is shown that a single-system Copenhagen measurement protocol is equivalent to 
the two-system von Neumann scheme with the memory filling up the system with negative information similar to the 
Dirac sea of negative energy. 
After equating the two quantum measurement protocols, we then apply this equivalence to the black hole radiation.  
That is, the black hole evaporation corresponds to the quantum measurement 
process and the two evaporation approaches, the observable-based single-system and the two-system entanglement-based protocols, can be made 
equivalent using quantum memory.  In particular, the measurement choice $\theta$ with the memory state inside the horizon in the entanglement-based scheme is 
shown to correspond to the observable of the measurement choice $\theta$ outside the horizon 
in the single-system protocol, that is, $\mathcal{O}_{\theta}^{out} = Q_{\theta}^{in}$. 
This indicates that the black hole as quantum memory is filling up with negative information outside the horizon, and its entropy 
corresponds to the logarithm of a number of equally probable measurement choices. 
This shows that the black hole radiation is no different than ordinary quantum theory.

\end{abstract}


\maketitle
One of the biggest differences between classical 
and quantum physics has been demonstrated using   
Bell's inequalities in entanglement \cite{bell}.  Moreover, 
with the recent rapid development of quantum information
technology, entanglement has been shown to have 
substantial importance \cite{nielsen}. 
On the other hand, interest in the so-called quantum 
measurement problem and quantum foundations have also resurged recently 
(for example, see \cite{mensky} and the references therein).  One of the puzzling 
features of quantum measurement, apart from its 
probabilistic nature, is that it has two quite different measurement protocols. 
While the single-system Copenhagen measurement protocol deals with only 
one system and observables, the two-system von Neumann protocol introduces  
the second system as a memory and uses entanglement. 
In this paper, using the negative entropy of entanglement \cite{cerf}, we will show the equivalence of these  
two quantum measurement schemes.  
The approaches can be made equivalent by 
considering that 
the second system of von Neumann's approach 
fills the system similar to Dirac's sea of negative energy 
filling up the vacuum.

We then apply the equivalence of the two quantum measurements to the process of black hole 
evaporation.  In particular, we wish to discuss two 
different approaches to black hole radiation, the observable-based single-system \cite{hawking1} 
and the entanglement-based two-system \cite{unruh1} protocols.  
 Similar to the measurement 
case, these two approaches yield the same thermal 
radiation of a system outside the black hole.  
We will suggest that the black hole radiation should be the quantum measurement process.  
In particular, we will argue that the single-system evaporation approach should 
correspond to the single-system Copenhagen measurement protocol 
and the entanglement-based approach should correspond to the two-system von Neumann 
measurement scheme.  Based on this, we discuss
the equivalence of these two approaches using negative entropy 
of entanglement. Therefore, the observable of the measurement choice $\theta$, $\mathcal{O}_{\theta}^{out}$, outside the horizon, in 
the single-system process is shown to be the same as the measurement choice made 
by the memory state, $Q_{\theta}^{in}$, inside the horizon in the two-system entanglement-based approach, that is,  
\begin{equation}
\mathcal{O}_{\theta}^{out} = Q_{\theta}^{in}
\label{TheEquation}\end{equation}
Moreover, it is suggested that the entropy of the black hole should correspond to the 
logarithm of a number of possible equally probable measurement choices, $\theta$, of the observer outside the horizon

Quantum theory has shown to yield many surprising and
counter-intuitive features that are unparalleled in classical theory \cite{nielsen}.  One
of these features includes the negative entropy for entangled
systems. In \cite{cerf}, it is shown that conditional entropy, which is the needed
information in order to have full knowledge about the other
part after the gain of shared information between the two
parts, can be negative in quantum theory.  
The negative entropy, with a 
quantum system, $S$, and a memory, $M$, is written as    
$H(S|M) = H(S:M) - H(M)$ where $H(S|M)$ denotes an 
amount of information about the system,  
 given the memory.
There have been a number of notable interpretations of 
negative entropy. 
For instance, negative entropy has been interpreted 
as potential information \cite{horodecki}
 and as reversible work \cite{vedral}. 
 In this paper, we will show that negative entropy is 
 a reasonable candidate to describe memory in 
a quantum measurement scheme.

Firstly, two simple classical procedures will be reviewed. 
Initially the particle is free to move around in a box.  
Therefore, initially no information 
is stored in the system, and the particle is at 
random, i.e., it could be either on the right 
or left side.  Then the partition is placed such that 
the particle is placed either on the left or 
right side of the partition. 
The randomness (in a sense that the particle 
can be either in the left or right) is now 
removed, and there is one bit of information. 
We now consider a reverse process.  
Initially, due to the partition, 
the particle is placed either in left or 
right side.  Therefore, there is 
no randomness as far as the particle's 
position (either left or right) 
is concerned.  Moreover, it has 
information of one bit stored.  
Later, the partition is removed, and the 
particle is free to move around. 
There is no information stored, and the 
uncertainty has emerged such that 
the particle can be either in the left or right side.

The concept of entropy is 
not only very interesting and useful but also 
puzzling.  
Firstly, entropy has been extremely powerful tool 
in thermodynamics both macroscopically and statistically.  
Moreover, it has also been very powerful 
in information theory, in that it 
has been used to quantify the amount of information. 
Interestingly, entropy quantifies two seemingly 
opposite entities, which therefore sometimes causes confusion, 
uncertainty (or ignorance) and information (or order). 
In order to illustrate the above discussion 
involving two processes  
in terms of entropy, we try to distinguish 
between these two types of entropy, 
where $H_{\it{ran}}$ denotes the entropy that 
quantifies uncertainty while 
$H_{\it{info}}$ describes the amount of information 
as used in information theory. 
Let us try 
to describe the process of learning or observing.  
In regards to the process of observing or 
learning for an observing party, the initial state 
should be that there is no information 
stored, and after observing the system, there is gained information.  
In terms of the entropy that quantifies the amount of 
information, initially, 
 $H_{\it{info}}=0$, and finally, $H_{\it{info}}=+1$.  
 Therefore, the change of 
 entropy can be considered as $\Delta H_{\it{info}}=+1$.  
 However, as discussed above, 
 the same process can be considered with the entropy that 
 quantifies uncertainty or randomness.  
 That is, initially $H_{\it{ran}}=+1$, and later $H_{\it{ran}}=0$.  
 Therefore, the change of entropy that quantifies 
 uncertainty corresponds to the negative value; 
 i.e., $\Delta H_{\it{ran}}=-1$.  Therefore, when the 
 learning process is described in terms 
 of the uncertainty entropy, it yields the negative value.  
 It is important to remember which entropy 
 (i.e., between information and uncertainty)
 that we are using when describing the system 
 $+$ memory. 
For example, if we are using the entropy of uncertainty for a system, 
then we should use the same type entropy for the memory 
as well (and similarly for entropy of information).  
That is, we should not use the entropy of uncertainty for 
the system and the entropy of information for the memory. 
We now apply this logic to the process of quantum measurements.


The quantum measurement scheme introduced by 
von Neumann involves a system and a memory 
(also called an apparatus).
With the interaction, the system and the 
memory become entangled and the measurement yields a probability distribution. 
Let us discuss the process of the system 
and the memory of the quantum measurement 
in terms of the two entropies, $H_{\it{info}}$ 
and $H_{\it{ran}}$, as discussed above. 
Initially, 
the system is in a pure state; therefore its 
entropy that quantifies uncertainty is zero; 
i.e., $H_{\it{ran}}^{\it{i}}=0$. After being 
entangled with the memory, the system 
is in a mixed state and the uncertainty 
of the system is quantified with the entropy as 
$H_{\it{ran}}^{\it{f}}=+1$.  Therefore,
the entropy of uncertainty has changed from $0$
 to $+1$, which can be written as 
$\Delta H_{\it{ran}}=+1$.  
On the other hand, the initial state of the memory 
may be considered after entanglement 
with the system.  Therefore, the uncertainty 
entropy of the initial state of the memory 
can be written as $H_{\it{ran}}^{\it{i}}=+1$.  
After the measurement, the state of the memory 
is in a pure state; therefore, $H_{\it{ran}}^{\it{f}}=0$. 
Therefore, the change of entropy may be written 
as $\Delta H_{\it{ran}}=-1$.  The von 
Neumann measurement can also be reviewed in terms of
the entropy that quantifies information, 
similarly to the Shannon's approach. 
Initially, 
the system is in a definite state; therefore, 
$H_{\it{info}}^{\it{i}}=+1$.  
After being entangled with the memory, the 
system is in  a random state and has no 
information; i.e., $H_{\it{info}}^{\it{f}}=0$. 
Therefore, the change of information entropy 
yields a negative value: $\Delta H_{\it{info}}=-1$. 
As before, the change of the information entropy 
is just the opposite compared to the uncertainty entropy. 
Similarly, the initial state of memory can be 
quantified in terms of information; i.e., initially, 
$H_{\it{info}}^{\it{i}}=0$. The final state of quantum 
memory is in a definite state; 
therefore, $H_{\it{info}}^{\it{f}}=+1$. 
Therefore, the change of information entropy may be written as 
$\Delta H_{\it{info}}=+1$.  

Therefore, it can be seen that when we take the entropy 
of uncertainty, the change of entropy for the system yields a positive 
value.  However, if we consider the change of 
uncertainty entropy for the memory, it yields a negative value, as 
discussed in \cite{cerf}. 
Therefore, the two-system entanglement-based von Neumann measurement model presents a natural 
realization of negative entropy. 

In \cite{cerf}, it has been shown that the value of 
quantum conditional entropy yields a negative value.  
Given a maximally entangled state, when one qubit is measured, 
the unknown value of the other qubit is $-1$.  
The qubit-antiqubit entanglement was then proposed, where the 
antiqubit is carrying $-1$ information, 
while the qubit is carrying $+1$ information. 
It is noted that the situation of 
negative entropy and antiqubits resembles that of negative 
energy and antiparticles in relativistic quantum theory. 
If we apply the interpretation of 
\cite{cerf} to the two-system quantum measurement, 
the system that is being measured is to be identified with 
a qubit, while the memory can be considered  as
the antiqubit associated with the Dirac sea of negative entropy,  
which is also consistent with our previous discussion of two-system measurement. 
When negative energy solutions were obtained, 
Dirac assumed the vacuum to be filled with  
negative energy to avoid an electron occupying 
the negative energy states. 
Dirac's interpretation of the vacuum as a 
filled negative energy sea was not merely an  
interpretation; indeed, it yielded a prediction of a 
positron with a hole in the sea.

\begin{figure}
\begin{center}
{\includegraphics[scale=.4]{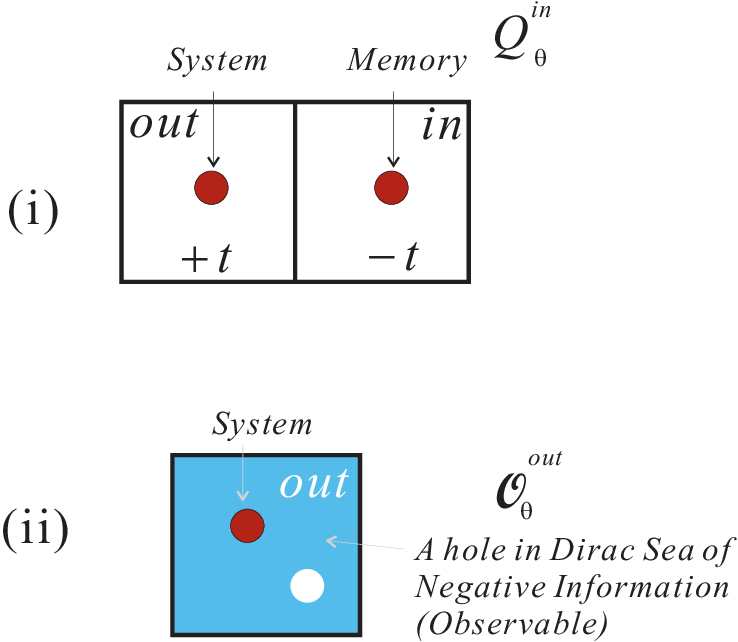}}

\end{center}
\caption{ (i) The two-system entanglement-based model where the second system is serving as memory 
(ii) Quantum memory: single-system model filled with negative information. 
The equivalence of (i) and (ii) occurs by equating the memory state $Q_{\theta}^{in}$ in (i) 
with the hole in the Dirac sea of negative information, or an observable $\mathcal{O}_{\theta}^{out}$
outside the horizon in the single-system protocol where $\theta$ corresponds to the choice of measurement.   }
\label{DiracSea}\end{figure}

We wish to apply logic similar to Dirac's 
interpretation to quantum memory in two-system quantum measurement. 
Given negative entropy, we assume the vacuum to be filled with negative 
information and, therefore, define the antiqubit to be a hole in the Dirac sea of negative information in the single-system, as in (ii) of Fig \ref{DiracSea}. 
That is, the two-system measurement can be considered as  
a single system that is filled with negative information,  
where the hole in the negative sea is acting as a frame of reference to measure the system, 
or the observable in a single-system Copenhagen measurement scheme.  
Let us discuss this equivalence between the two-system 
von Neumann protocol and the single-system Copenhagen scheme with black hole radiation.

%

In \cite{hawking1}, it was shown that black hole radiation was caused by 
a gravitational collapse and that the outgoing 
radiation is independent of the collapsing star.  
The particle radiation was derived with 
the observable of the number operator using Bogoliubov transformation. 
Just as in the case of the Copenhagen single-system measurement, 
it has the state and observable of just the outside of the black hole. 
That is, the protocol of black hole radiation may be considered 
as a single-system process with the state and the observable of the single
system. 
On the other hand, another method based on entanglement between two systems was introduced \cite{unruh1}  
to derive black hole evaporation by considering 
an eternal black hole.  
Following \cite{maldacena}, let us consider the following entanglement-based microcanonical state for 
two systems, that is, outside and inside the horizon,  
\begin{equation}
|\psi\rangle = \frac{1}{\sqrt{\Omega}}\sum_{\theta =0}^{\Omega-1} |\theta\rangle_{out} |\theta\rangle_{in}
\label{Micro}\end{equation}
where $\Omega$ corresponds to the number of equally accessible states of a black hole. 
Therefore, just as in the measurement schemes, 
the single-system and two-system approaches are quite 
different since the latter requires entanglement 
in two systems while the former does not.

When the black hole was first shown to have entropy \cite{bekenstein} and radiation \cite{hawking1} as a thermal object, 
one of the central issues 
was its information loss process (see \cite{paradox} and the references therein), that is, it seemed to provide  
non-unitarity (as can be seen when we trace out inside the horizon in (\ref{Micro})) 
and many considered it to be a serious threat to unitarity in quantum theory.    
However, quantum theory includes not only unitary evolution but also non-unitary process as 
well, that is, the measurement.   Therefore, if we assume quantum theory 
(both unitary evolution and measurement) is correct, and black hole evaporation is 
indeed non-unitary, then we should come to the conclusion that the radiation must be the measurement process!
Interestingly, similar to the quantum measurement case, there are 
two types of black hole radiation, the observable-based single-system and the entanglement-based two-system protocols. 
We wish to construct black holes as quantum memory based on the equivalence 
of the entanglement-based protocol and the single-system scheme in terms of degeneracy.

Previously, we discussed the equivalence of the single-system 
Copenhagen protocol and the two-system von Neumann protocol with negative information 
filling the system.  
Similarly, as shown in Fig. \ref{DiracSea}, 
the memory system inside the horizon in the entangled state (\ref{Micro}) can be considered a Dirac sea of negative information 
filling the system outside the horizon, which is therefore equivalent to the single-system radiation. 
Since the states $|0\rangle, \cdots, |\Omega-1\rangle$ in (\ref{Micro}) 
are equally accessible states for the black hole, there is an observable $\mathcal{O}_0$ such that 
these degenerate states yield the same eigenvalue with respect to this observable.   
Due to the symmetry between the Schr\"odinger and Heisenberg pictures of
quantum theory, we may also establish, locally distinctive, degenerate observables, 
$\mathcal{O}_0, \cdots, \mathcal{O}_{\Omega-1}$, given the state $|0\rangle$.  
For instance, let us consider the unitary transformation of the accessible states as follows,   
$\mathcal{U}_{\theta}|0\rangle = |\theta\rangle$ where $0\leq \theta\leq \Omega-1$.
 With this unitary transformation, we may establish, locally distinctive \cite{local}, degenerate observables as follows, 
$\mathcal{O}_{\theta} \equiv \mathcal{U}_{\theta}^{\dagger} \mathcal{O}_0  \mathcal{U}_{\theta}$ such that  
\begin{equation}
_{out}\langle 0| \mathcal{O}_{\theta}^{out} |0\rangle_{out}, 
\label{Observable}\end{equation} 
yields the same value for all $0\leq \theta \leq\Omega-1$.
With these observables, we may now equate the entanglement-based two-system protocol in (\ref{Micro}) to a single-system protocol, 
as seen with the Copenhagen measurement protocol. 
That is, each memory state $|\theta\rangle_{in}$ in (\ref{Micro}) (or $Q_{\theta}^{in}$ in (ii) of Fig. \ref{DiracSea}) 
should correspond to the observable $\mathcal{O}_{\theta}^{out}$ in the single-system 
protocol in (\ref{Observable}) (or (i) of Fig. \ref{DiracSea}).  
  (Note that in (\ref{Observable}), it has both the state and the observable of just the system outside the horizon.)

Moreover, in the case of (\ref{Micro}), $|\psi\rangle$ describes the system outside and the memory inside the horizon where each 
$|\theta\rangle_{in}$ corresponds to the choice 
of measurement (i.e., $\mathcal{O}_{\theta}^{out}$ in the single-system protocol) 
and there are $\Omega$ equally probable choices of measuring the system $|0\rangle_{out}$ with 
the memory inside the horizon.  
Therefore, the black hole is serving as quantum memory in measuring the system outside the black hole by filling up the outside 
with negative information. 
Moreover, $\Omega$ corresponds to the number of $\theta$'s, which are equally 
probable ways of choosing the memory in the microcanonical state (\ref{Micro}) with entropy,  
$S=k\ln \Omega$, where $k$ is the Boltzmann constant. 
That is, the black hole entropy yields a number of ways the observer outside the horizon  
may choose to measure the black hole. 
Therefore, this shows the equivalence between the two 
different schemes in black hole radiation, such that the  
$\Omega$ equally probable measurement choices in the two-system 
protocol are equivalent to the 
$\Omega$ equally probable choices of 
observables in the single-system scheme. 
In other words, the choice $\theta$ with 
memory inside the horizon in (\ref{Micro}) corresponds to the observable choice 
$\theta$ of the system outside in (\ref{Observable}), as seen in (\ref{TheEquation}).

We wish to discuss some of the points that may be dealt with by treating black holes as quantum memory. 
First, one of the puzzles involving black hole radiation is its 
information loss problem, that is, the initial pure state that evolved into the 
final mixed state.  By treating this non-unitary process as a quantum measurement, 
black hole radiation may be considered one of the two ordinary quantum 
mechanical processes.

Second, quantum memory yields a natural interpretation of the negative entropy 
shown in quantum entanglement.  Not only does negative entropy provide 
a natural interpretation of the measurement process in the two-system von Neumann scheme, it makes the two-system 
model equivalent to the single-system observable-based protocol by providing a type of Dirac sea of negative information.
\\
{\it{Acknowledgments}}: The author is grateful for hospitality of Chulalongkorn University while part of this work was carried out.


\begin{thebibliography}{0}
\bibitem{local} Although the observables as a whole will not change, 
the observables will be different locally and the way to measure with these degenerate observables will be different.  
For example, in the case of two qubits, $|00\rangle$ and $|11\rangle$ (or, $|01\rangle$ and $|10\rangle$) 
are degenerate states with respect to the observable $\sigma_z \otimes \sigma_z$. 
We may also establish degenerate observables, $\sigma_z \otimes \sigma_z$ and $(-\sigma_z)\otimes (-\sigma_z)$, with respect to the state $|00\rangle$.
Although these two observables are the same, they are different locally.  Moreover, the way each observable is applied 
during the measurement will be different (think of Stern-Gerlach experiment, for instance).
\end{thebibliography}
\end{document}